# Defect-trapped hydrogen couples stress-triggered atomic decohesion with explosive crack growth

L. Gao[1,2]*†, T. Schwarz-Selinger[2], M. Balden[2], C. Li[3], P. Manz[4], W. Jacob[2], R. Neu[2], Ch. Linsmeier[1], J. Wang[5]* and G.-H. Lu[6]*

[1]Forschungszentrum Jülich GmbH, Institute of Fusion energy and Nuclear waste management – Plasmaphysik (IFN-1), 52428, Jülich, Germany;
[2]Max-Planck-Institut für Plasmaphysik (IPP), 85748, Garching, Germany;
[3]Shanghai Institute of Applied Physics, Chinese Academy of Science, 201800, Shanghai, China;
[4]University of Greifswald, Institute of Physics, 17489, Greifswald, Germany;
[5]Department of Mechanical and Materials Engineering, University of Nebraska, 68588, Lincoln, USA;
[6]Beihang University, School of Physics, 100191, Beijing, China.

*Corresponding authors: li.gao@extern.fz-juelich.de (L. Gao), jianwang@unl.edu (J. Wang), lgh@buaa.edu.cn (G.-H. Lu)
†Present address: Gauss Fusion GmbH, Garching, 85748, Germany

**Hydrogen embrittlement (HE) has remained a major challenge for over a century[1]. The prevailing view that *diffusive hydrogen* governs embrittlement has fostered the assumption that defect trapping mitigates HE[2-7]. Here, we challenge this paradigm by showing that *defect-trapped hydrogen* enables internal cleavage fracture under sub-threshold stress. Sequential hydrogen-isotope-plasma exposure and tungsten-ion irradiation distinguish defect decoration and mechanical triggering from the subsequent cavity growth, revealing a two-stage mechanochemical pathway for fracture initiation. First, hydrogen accumulation at dislocation cores lowers the local decohesion threshold, rendering the reconstructed core susceptible to a small external mechanical trigger for atomic bond rupture. This rupture nucleates an internal free surface and marks the onset of crack formation. We propose that hydrogen recombination could release chemical energy within the nascent crack and drive a rapid brittle growth into a mesoscopic cavity. By separating the mechanical trigger for atomic decohesion from the proposed energetic driving force for subsequent crack growth, our results establish an atomistic foundation for the classical hydrogen-enhanced decohesion model[8-11] and a mechanistic framework for hydrogen-induced crack formation. Finally, shifting the focus from experimentally elusive diffusive hydrogen to accessible trapped hydrogen[12-14] identifies the local trapped-hydrogen state as a potentially quantifiable control parameter for hydrogen-induced fracture.**

Hydrogen embrittlement (HE)[2-7] is a hydrogen-assisted fracture phenomenon that can severely compromise the mechanical integrity of metals. Extensive efforts to resolve the underlying physics of hydrogen-induced failure have largely focused on hydrogen effects at established crack tips or on deformation-mediated processes, invoking mechanisms including hydrogen-enhanced decohesion (HEDE[8-11]), hydrogen-enhanced localized plasticity (HELP[15-17]), adsorption-induced dislocation emission (AIDE[18]), hydrogen-enhanced vacancy formation[19], and hydride formation. These mechanisms have provided important insights into hydrogen-assisted degradation and crack evolution, but they do not directly resolve the atomistic event that converts hydrogen–defect interactions[20] into the first irreversible crack.

A central question is whether primary crack nucleation requires hydrogen to remain mobile during deformation. Experiments on hydrogen-charged polycrystalline nickel[21] have shown that intergranular microcracking can still occur at cryogenic temperature, where deformation-induced hydrogen redistribution is strongly suppressed. Establishing that mobile H–deformation interactions are not an intrinsic requirement for hydrogen-induced intergranular fracture, these observations suggest that hydrogen segregated to grain boundaries before deformation may dominate crack initiation under the investigated conditions. However, substantial mechanical deformation is still applied to initiate these intergranular cracks in these experiments, and the hydrogen population directly responsible for the initiating event was not isolated or quantified at the atomic scale. Whether primary crack nucleation can occur without concurrent plastic deformation or dislocation motion remains unresolved.

Hydrogen trapped at defects provides a natural route to address this question. Such defect-trapped hydrogen (DTH) is commonly regarded as a reservoir that removes hydrogen from the mobile population and can therefore mitigate embrittlement. However, trapping also concentrates hydrogen at specific atomic environments where local chemical bonding, lattice distortion, and defect energetics can be profoundly altered. This raises a fundamental possibility that DTH itself may alter the mechanical stability of a defect sufficiently to permit primary crack nucleation under a small external mechanical perturbation. More specifically, can DTH function as an active state variable that directly controls the onset of atomic decohesion, rather than merely serving as a precursor to deformation-assisted fracture? Addressing this question experimentally is challenging because hydrogen transport, plastic deformation, and defect evolution are normally coupled during conventional HE experiments. Under such conditions, it is difficult to determine whether fracture is triggered by hydrogen-mediated modification of deformation processes, by direct destabilization of hydrogen-decorated defects, or by a synergy between these effects.

Resolving this question requires an experimental configuration that separates hydrogen decoration from mechanical triggering and deformation-mediated relaxation. In particular, the applied stress must be sufficient to trigger the local fracture instability, while remaining below the threshold for dislocation glide and substantial plastic relaxation. Here we implement this separation through a sequential deuterium-plasma exposure and tungsten-ion irradiation of recrystallized body-centered-cubic tungsten[22] (bcc W). Tungsten as a non-hydride-forming metal provides a comparatively well-defined environment for interrogating hydrogen–defect interactions without the additional compositional and phase complexities of engineering alloys like steels. In the following, we use the term '*hydrogen*' to describe the behavior of hydrogen isotopes in general; 'deuterium (D)' will be utilized only for specific experimental description. As illustrated in Fig. 1, D is first introduced under low-fluence conditions[23,24] to decorate pre-existing defects and establish accessible DTH states. The material is then subjected to MeV W-ion irradiation to generate a controlled mechanical perturbation[25-27] and internal tensile stresses for fracture initiation in depth beyond the W-ion implantation zone (W-IPZ), while maintaining stresses well below the critical resolved shear stress for dislocation glide in W. Importantly, no concurrent hydrogen supply is provided during this stage, thereby avoiding the coupled hydrogen accumulation and cavity inflation that accompany continuous plasma exposure and hydrogen blistering. Following fracture initiation, a second, gentler D-plasma exposure is used to further inflate the pre-existing cracks with sub-microscopic heights to dimensions accessible by microscopy. This experimental design separates the pre-fracture defect-decoration and irradiation-triggered fracture stages from subsequent plasma-assisted cavity inflation, while avoiding concurrent hydrogen supply during the mechanical triggering stage.

***Table 1 HIC formation in W samples with different treatments during sequential irradiations.*** *Different samples (i.e., W2-W5) and varied surface areas on one sample (i.e., A1-A4 on W1, see Fig. 1) are prepared according to their irradiation histories. Samples W2-W5 were made to create lower D occupancies at defects compared with that at A4 in W1 via either elevating temperature in the 1st D plasma loading (i.e., sample W2&3) or post-decoration annealing (i.e., sample W4&5). None of them shows HIC formation after the same W ion irradiation and 2nd D plasma inflation, demonstrating the sensitive dependence of HIC nucleation on the local D occupancy.*

| Treatments / Samples | 1st D plasma exposure @ $T_1$ | Annealing @ $T_{ann}$ | 5 MeV $W^{3+}$ ion irradiation | 2nd D plasma exposure @370K | HIC areal density ($mm^{-2}$) |
|---|---|---|---|---|---|
| W1_A1 | × | × | × | ✓ | 0 |
| W1_A2 | ✓ ($T_1$=300K) | × | × | ✓ | 0 |
| W1_A3 | × | × | ✓ | ✓ | 0 |
| W1_A4 | ✓ ($T_1$=300K) | × | ✓ | ✓ | **57.8** |
| W2 | ✓ (***$T_1$=370 K***) | × | ✓ | ✓ | 0 |
| W3 | ✓ (***$T_1$=450 K***) | × | ✓ | ✓ | 0 |
| W4 | ✓ ($T_1$=300K) | ✓ (***$T_{ann}$=370 K***) | ✓ | ✓ | 0 |
| W5 | ✓ ($T_1$=300K) | ✓ (***$T_{ann}$=450 K***) | ✓ | ✓ | 0 |

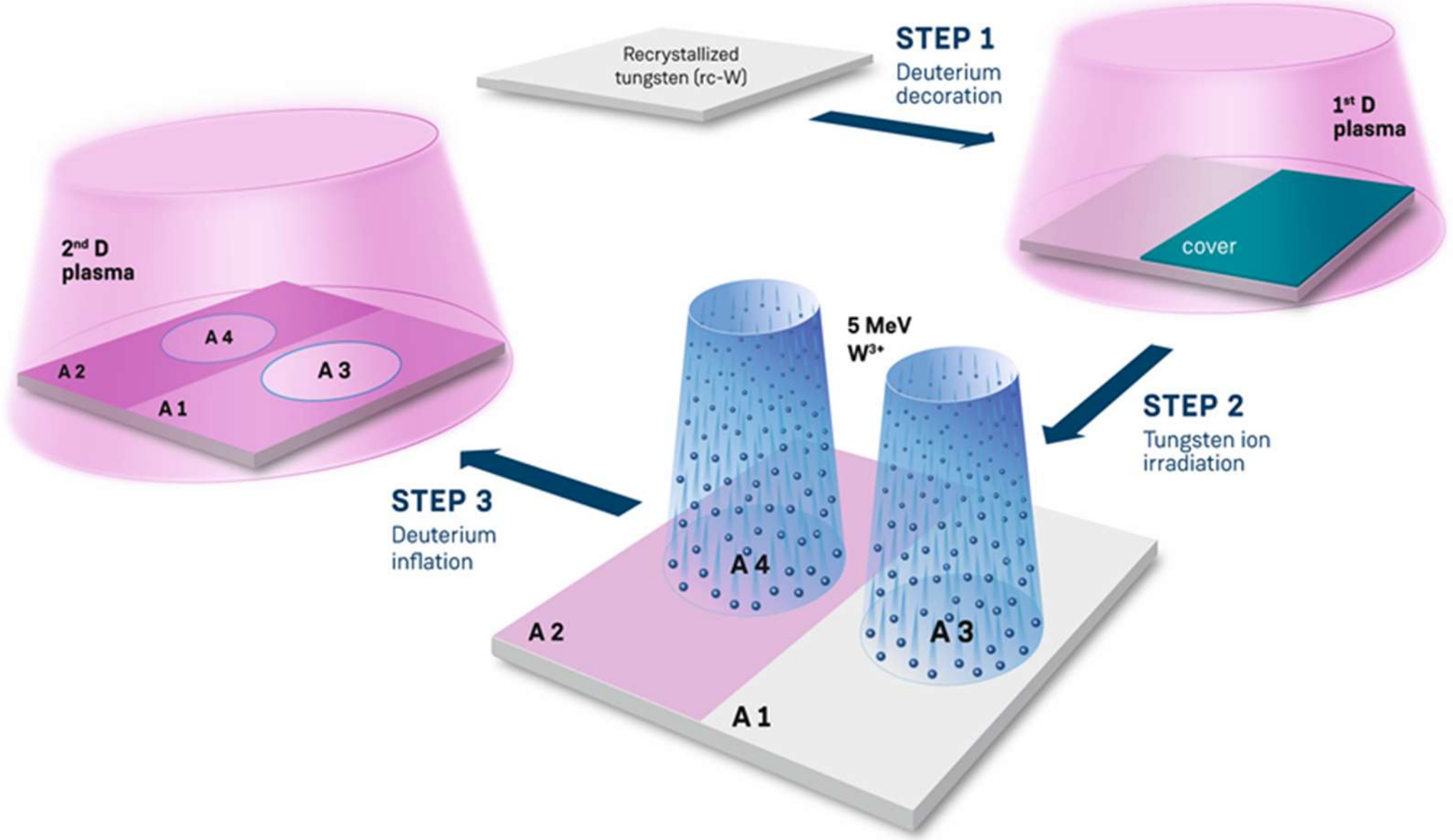

***Figure 1. Sequential exposure strategy for subsurface crack formation in tungsten.*** *Four areas (A1–A4) of a recrystallized W sample (W1) received different combinations of 1st D-plasma exposure, W-ion irradiation and 2nd D-plasma exposure. HICs were observed only in W1_A4, which received all three treatments. The A4–A3 comparison highlights the role of pre-fracture D decoration.*

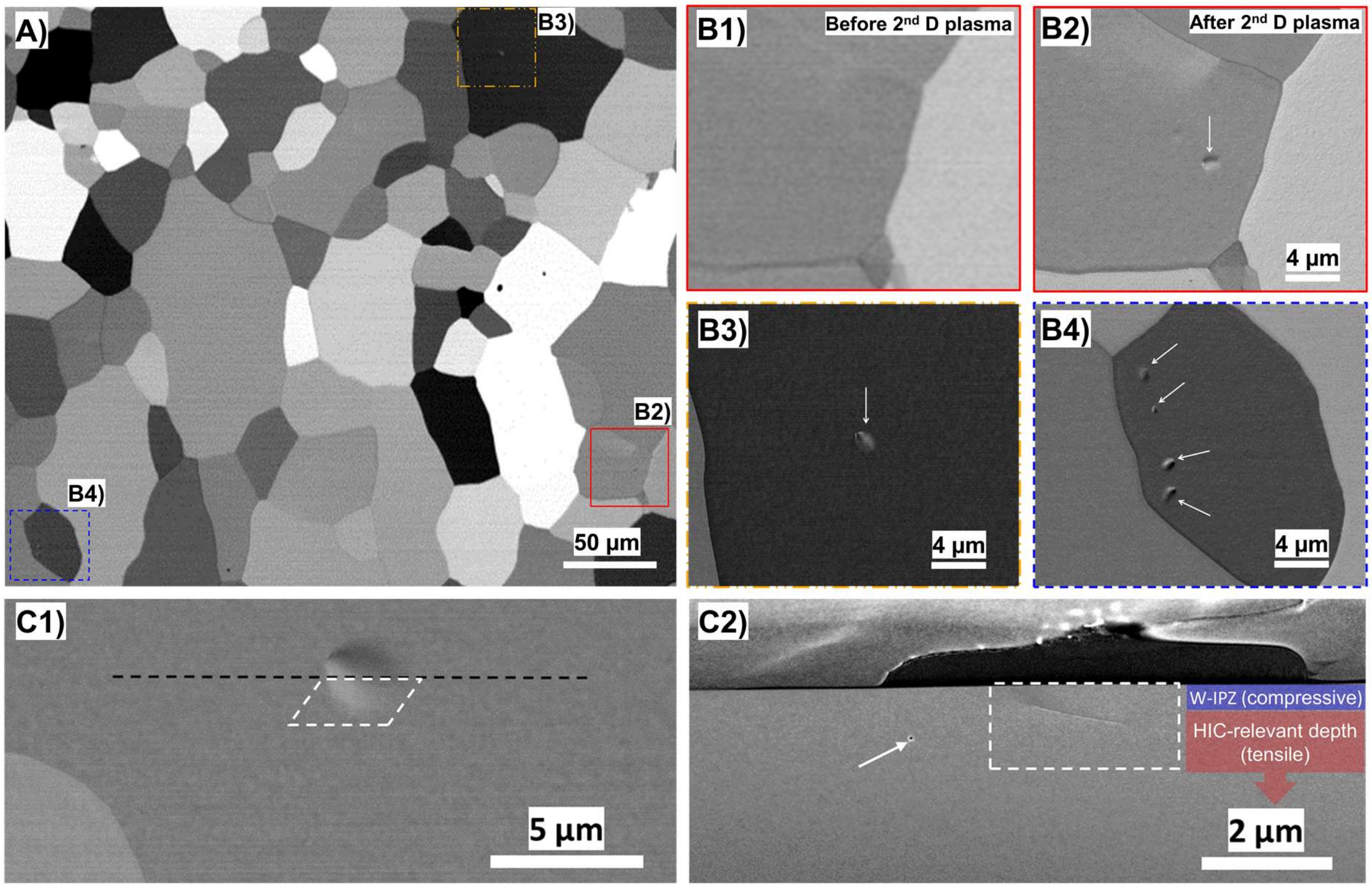

***Figure 2 | Subsurface HICs and their subsequent inflation.*** *A) HIC-forming area W1_A4 after the second D-plasma exposure. B1–B2) The same location before and after the second exposure; B3–B4) Additional intragranular HICs, with surface blisters indicated by arrows. C1) Tilted surface view of an HIC-associated feature; C2) FIB cross-section of the HIC in C1 showing the subsurface cavity (dashed box) and a separate nanopore (arrow). The cavity midpoint is approximately 850 nm below the original surface, compared with the estimated maximum W-ion implantation depth of approximately 750 nm. This comparison is consistent with, but does not prove, nucleation beyond the implantation zone. B1 is a lower-resolution, zoomed-in view of the same location before the second exposure.*

Four additional W samples (i.e., W2-W5, see Table 1) were heated prior to W ion irradiation to acquire lower local D occupancy levels at intrinsic defects compared with W1_A4 (see Fig. 1). HIC-associated blisters were observed only in W1_A4, which received all three sequential treatments (Figs. 1 and 2). No HICs were detected in the other W1 areas or in samples W2–W5, which had lower D occupancy following elevated-temperature exposure or annealing (Table 1 and Fig. S1). The observation that HICs form exclusively on W1_A4 (Fig. 2) is statistically confirmed by SEM measurements on sufficiently large areas (*Fig. S1, see Supplementary Note 1*). Comparing A4 and A3 (with or without the 1$^{st}$ D plasma exposure for pre-fracture D decoration) clearly demonstrates that pre-existing DTH is indispensable for HIC nucleation in the absence of either appreciable diffusive hydrogen (*see Supplementary Note 2*) or significant plastic deformation. While comparing A4 with A2 shows that DTH alone is insufficient: HIC nucleation additionally required the mechanical perturbation introduced by W-ion irradiation. The complete suppression of HIC nucleation in samples W2–W5 with slightly lower DTH occupancy (see Table 1) further reveals an occupancy-sensitive fracture initiation. These comparisons indicate that DTH establishes local susceptibility to fracture, whereas the sub-threshold stress associated with W-ion irradiation triggers instability.

Having established that DTH and external perturbation are both required for HIC nucleation, we next ask how DTH alters the mechanical response of the intrinsic defects such that this otherwise weak perturbation becomes sufficient to initiate primary crack nucleation. We focus on screw dislocations, which are primary carriers of plasticity in bcc metals and plausible nucleation sites for HICs (see justification in *Supplementary Note 3*) in bcc metals[11,28-31]. In their pristine state, the purely shear character of the stress field of a screw dislocation precludes direct coupling with the normal ($\sigma_N$) or hydrostatic components of an externally applied stress ($\sigma_a$). Upon the 1st D plasma exposure, these lower-electron-density core regions preferentially sequester dissolved hydrogen (see, e.g., Fig. 4A), because hydrogen behaves effectively as an electron acceptor in the W matrix, forming an anionic $H^-$ species[32]. Dissolved H atoms therefore experience strong exclusion from both the surrounding metallic electron sea and neighboring H atoms. As DTH accumulates at screw cores, the resulting H-H and H-W repulsive interactions induce localized lattice dilation and trigger a profound, symmetry-breaking core reconstruction[29,30,33].

Crucially, this structural transformation fundamentally alters the mechanical response of the dislocation core to $\sigma_a$. Rather than redistributing the resolved stress components, the induced volumetric dilation enables the reconstructed core to physically couple with the resolved normal stress component $\sigma_N$—an interaction strictly forbidden in the pristine state. The coupled normal stress ($\sigma_\perp$) begins to act on the screw core. Continued DTH accumulation further destabilizes the core by drastically lowering the critical normal stress threshold ($\sigma_C$) required for atomic bond rupture. Concurrently, the reconstructed core develops an elevated Peierls barrier[29-31,33] suppressing dislocation glide. Thus, DTH occupancy at the dislocation core orchestrates a dual embrittlement mechanism: it pins the dislocation to induce local hardening[34] while simultaneously promoting bond rupture via emergent normal-stress coupling ($\sigma_\perp > 0$) and the collapse of the decohesion barrier.

We define the atomic decohesion under an experimentally small external load as Stage I of the HIC formation process. Here, atomic decohesion denotes the loss of atomic cohesion leading to the nucleation of an internal free surface, marking the onset of internal crack surface formation. To visualize this instability governed by the DTH-occupancy, we replace the conventional temperature abscissa with DTH occupancy and construct in Fig. 3 a schematic stress–DTH phase diagram showing the collapsing drop in critical decohesion stress ($\sigma_C \rightarrow 0$) together with the increasing normal stress coupling factor ($\sigma_\perp/\sigma_N \rightarrow 1$). Notably, Stage I is proposed to be governed by a regime of localized *dynamic saturation*, in which the initially available trap sites become occupied while further core reconstruction generates additional sites for DTH accumulation. We define a relative occupancy $\theta$ – the ratio between the absolute number of DTH atoms and initially available trap sites at the screw segment core – to describe local DTH trapping; $\theta = 1$ denotes local saturation of the defined trap-site population. This explains the further rise of absolute DTH amount at the screw segment core while $\theta$ remains unity (see top X-axis in Fig. 3). This progressive degradation ultimately drives the system into a state where the normal stress coupled to the reconstructed screw core $\sigma_\perp$ from the applied stress $\sigma_a$ (i.e., $\sigma_a \approx 157$ MPa as the estimated W-ion irradiation-induced tensile stress, see *Supplementary Note 4*) becomes sufficient to trigger the atomic decohesion (i.e., $\sigma_\perp > \sigma_C$, see W1_A4 in Fig. 3). In contrast, samples with slightly lower DTH occupancy (W2–W5) or without DTH occupancy ($\theta = 0$, W1_A3) remain below the decohesion boundary, revealing a sharp bifurcation in fracture behavior upon reaching local dynamic core saturation. Importantly, since $\sigma_a$ via W ion irradiation and accordingly the resolved shear stress component remains far below the Peierls stress of screw dislocations in W at 300 K (~4 GPa[35]), conventional stress-driven dislocation glide is expected to be strongly suppressed under the applied conditions.

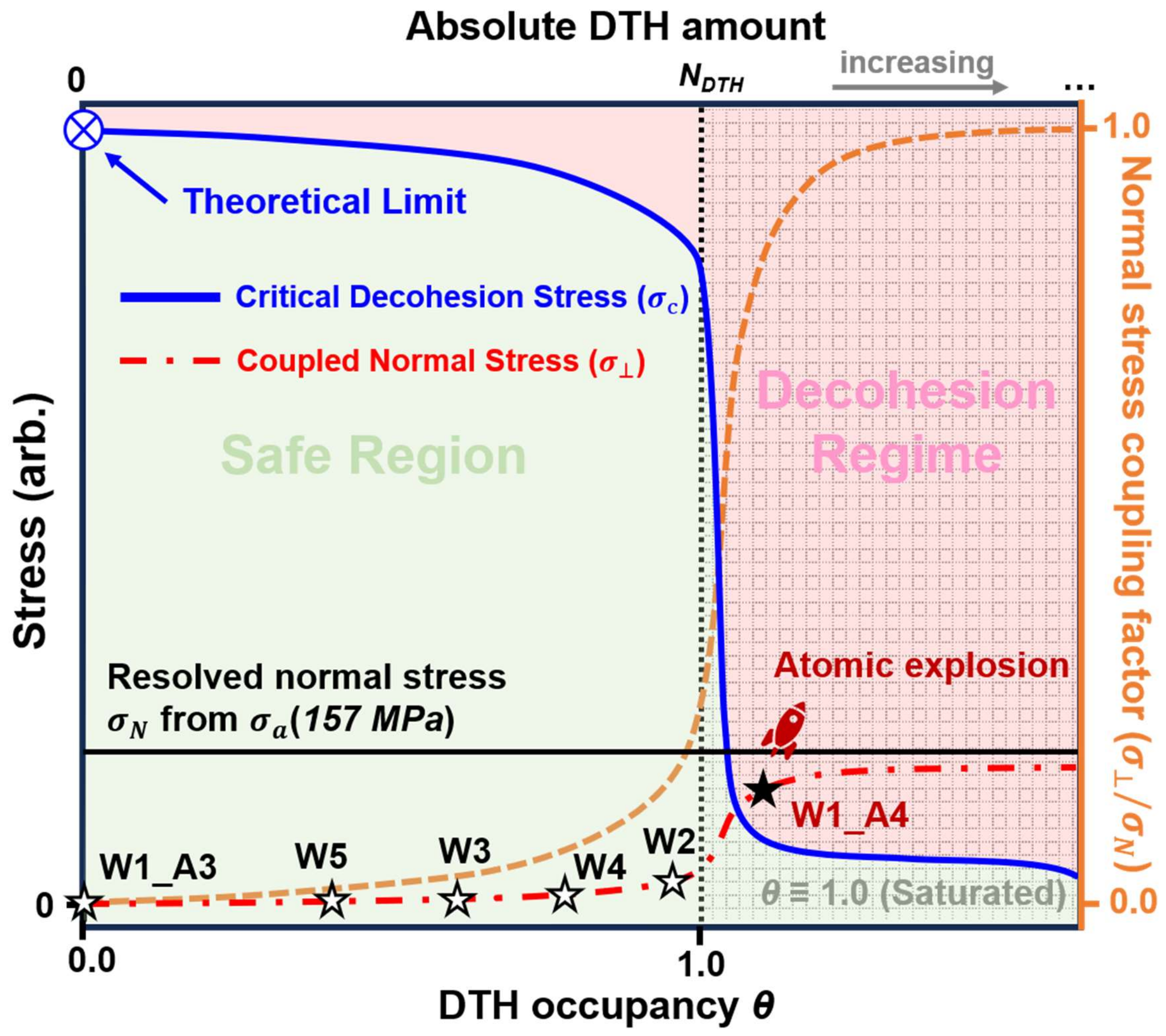

***Figure 3 Schematic stress–DTH phase diagram for Stage I atomic decohesion.*** *The blue solid curve represents the proposed decrease in critical decohesion stress ($\sigma_C$) with increasing DTH occupancy; the orange dashed curve represents the increasing coupling of the screw segment to the resolved normal stress ($\sigma_\perp/\sigma_N$). Decohesion is predicted when ($\sigma_\perp > \sigma_C$). The vertical dashed line marks the onset of dynamic DTH saturation ($\theta$=1); further DTH accumulation can increase core expansion without increasing the defined relative occupancy. The black star marks W1_A4 under the estimated tensile stress of approximately 157 MPa. W1_A3 and W2–W5 remain in the schematic safe regime. The "atomic explosion" symbol denotes the proposed subsequent Stage II process (Fig. 4).*

However, the subsequent formation of a primary cavity several hundred nanometers[36] in diameter presents a fundamental mechanical paradox with respect to the atomic-scale nascent crack. Even assuming full coupling of $\sigma_N$ into $\sigma_\perp$ ($\sigma_\perp / \sigma_N \approx 1$), it remains less than 1% of the theoretical cleavage strength of defect-free tungsten (~29 GPa[37]) and the lower cleavage strength at the pristine screw dislocation core[31]. This large disparity motivates the question of how an atomic-scale crack embryo could develop into the observed mesoscopic cavity under the present conditions. To resolve this discrepancy, we propose in Fig. 4 a two-stage mechanism for internal HIC formation. Following Stage I towards atomic decohesion described in Fig. 3, the newly exposed W atoms carrying dangling bonds become highly reactive electronic sinks. This abrupt local electronic redistribution destabilizes nearby DTH atoms, inducing an instantaneous electronic transition from anionic hydrogen ($H^-/D^-$) to neutral hydrogen ($H^0/D^0$). Adjacent neutral hydrogen/deuterium atoms subsequently recombine to form the first molecule ($2H^0 \rightarrow H_2$ / $2D^0 \rightarrow D_2$, Fig. 4B) at the newly formed crack surface.

The formation of a single $H_2/D_2$ molecule releases approximately 4.5 eV of recombination energy. Within the atomically confined nascent crack, such infinitesimal chemical energy release corresponds to a nominal energy-density scale of approximately 25 GPa under the geometric assumptions (see *Supplementary Note 5*). The extent to which this chemical energy is converted into mechanical work, and its quantitative contribution to crack propagation, remain to be established. The formation of internal mesoscale cavities (Fig. 2) demonstrates that substantial crack growth occurs under the present experimental conditions without an additional externally applied macroscopic load. We therefore propose that chemical energy released by deuterium recombination could contribute to the localized mechanical work associated with rapid crack growth along the energetically favored {100} cleavage planes[38]. We refer to this proposed transition from atomic decohesion to rapid crack growth as an "atomic explosion". This term denotes the hypothesized contribution of recombination-associated energy release to subsequent crack growth. The characteristic dislocation-free zones observed above the primary cavities[36] are consistent with predominantly decohesive growth, although they do not by themselves establish the absence of all dislocation-mediated plastic relaxation. Subsequent $2^{nd}$ plasma exposure (Fig. 1) then enables further gas accumulation[39,40] and continued cavity inflation, ultimately producing the SEM-detectable surface blistering (Fig. 2).

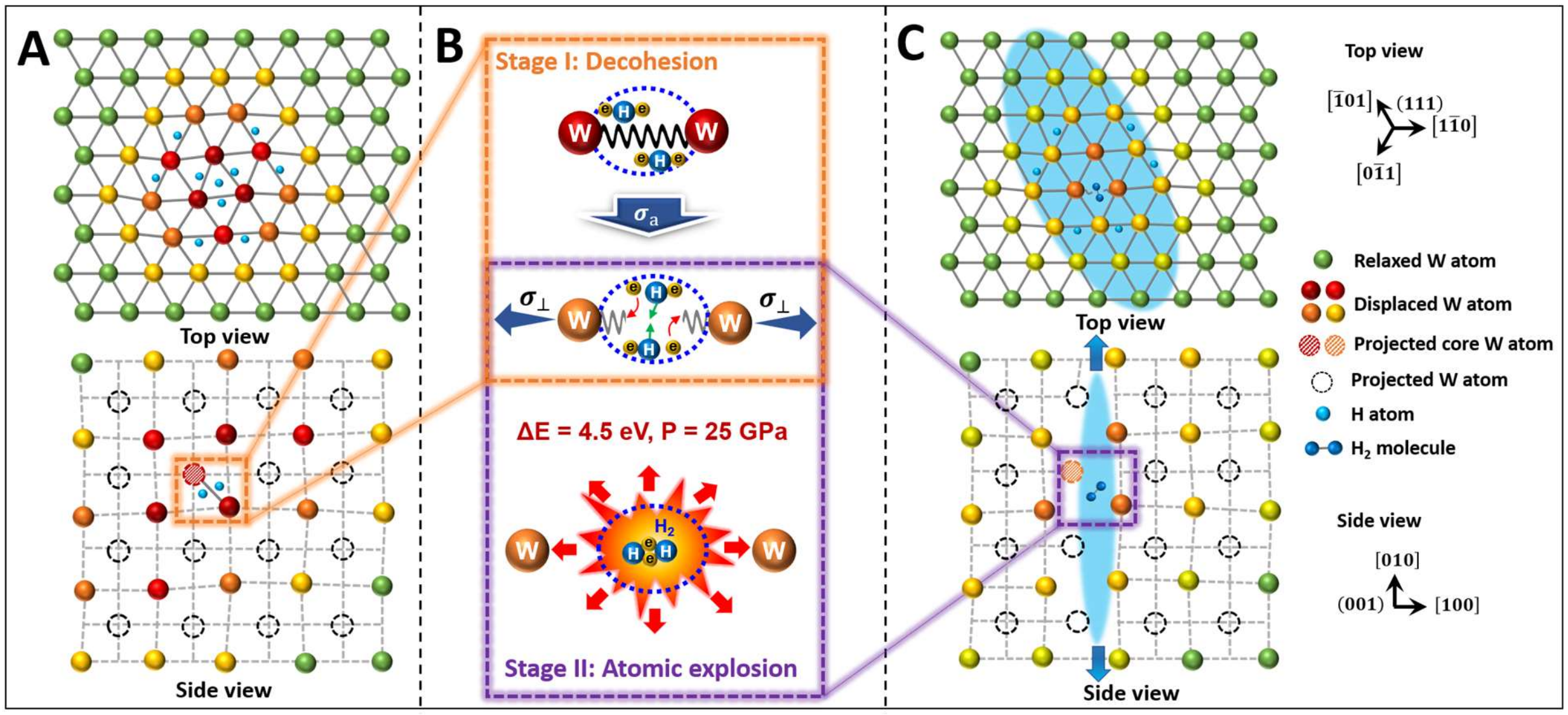

***Figure 4 Proposed two-stage mechanism of DTH-assisted internal fracture.*** *A) DTH accumulation induces screw-core reconstruction and lattice dilation, enabling coupling to the applied normal stress. B) Stage I: local atomic decohesion nucleates an internal free surface when the coupled stress exceeds the reduced threshold. Stage II: recombination of nearby hydrogen/deuterium atoms may release chemical energy and contribute to rapid crack growth along a {100} cleavage plane. C) Schematic evolution of the crack embryo into a mesoscopic cavity, consistent with predominantly decohesive growth without evident dislocation-mediated plastic relaxation. The light-blue region indicates the proposed cavity expansion associated with Stage II.*

Beyond its role as a mechanical trigger, W-ion irradiation offers an experimental advantage by allowing the irradiation-induced perturbation to be spatially confined to selected regions of a specimen. This enables within-specimen comparisons between regions with different treatment histories, while minimizing specimen-to-specimen variability in material microstructure and preparation. A constraint, however, is that HIC nucleation appears to be favored at screw-dislocation segments beyond the direct W-ion penetration zone (W-IPZ; ~750 nm; Fig. 2C and Fig. S2). Within the W-IPZ with compressive stress, continuous ion bombardment is expected to impede stable retention of the critical DTH configurations required for atomic decohesion (Fig. 3; see *Supplementary Note 6*). The exceptionally small estimated tensile stress (~157 MPa), compared with the theoretical cleavage strength of defect-free W, strongly supports a substantial DTH-induced reduction in the stress required for fracture initiation. Previous observations of HIC formation during deuterium-plasma exposure alone[23,36], without the W-ion-induced mechanical perturbation used here, provide further phenomenological support for the possibility that defect-trapped hydrogen can promote fracture under very small mechanical driving forces.

The rarity of HIC formation further solidifies the proposed mechanism. In W1_A4, the HIC number density is approximately three orders of magnitude lower than the screw-dislocation density in recrystallized W[22] (*see Supplementary Note 7*). Quantitative analysis gives a DTH line density of 3.5–26.3 × $10^5$ D atoms per meter of dislocation line, corresponding to an average occupancy of approximately $10^{-4}$ of the initially available sites along the dislocation cores (*Supplementary Note 8*). Although far below the occupancy associated with atomic decohesion, this average does not exclude local saturation at a small subset of screw segments, given the potentially non-uniform distribution of retained D. Such local enrichment remains a central hypothesis of the proposed rare-event mechanism. Only a minute fraction of screw segments may reach the critical configuration in which DTH-induced core reconstruction enables the coupled normal stress ($\sigma_{\perp}$) to exceed the local decohesion threshold ($\sigma_C$). The complete suppression of HIC formation after mild annealing (e.g., 370 K) is consistent with a reduced probability of forming these critical configurations, even when appreciable D remains trapped in the material.

Our findings provide an atomistic perspective on hydrogen-enhanced decohesion (HEDE[8-11]). In the proposed mechanism, DTH promotes HEDE not simply through hydrogen segregation to an existing fracture interface, but by reconstructing a screw-dislocation core and enabling the applied normal stress to participate in W–W bond rupture. This identifies a defect-mediated route to HEDE in which the local trapped-hydrogen configuration governs the onset of atomic decohesion. The primary cracks already undergo substantial opening along the cleavage plane before the second plasma exposure (Fig. 2-B1), indicating that their initiation cannot be attributed to subsequent mobile-hydrogen transport or spontaneous DTH-induced fracture. Instead, our results support a mechanism in which a highly D-decorated defect undergoes atomic decohesion under a small mechanical perturbation, without clear evidence of substantial dislocation-mediated plastic relaxation. The resulting internal free surface may provide a confined environment for deuterium recombination, whose associated chemical energy release could contribute to subsequent crack growth and cavity formation. The proposed mechanism may also explain the incubation period preceding delayed hydrogen-induced fracture. Rather than reflecting solely the diffusion of mobile hydrogen towards highly stressed regions, this period may represent the progressive accumulation of DTH at susceptible defects until a critical local configuration is reached. Although not directly resolved here, this interpretation suggests that defect decoration and local hydrogen accumulation may govern fracture initiation, rather than hydrogen diffusion alone.

These findings shift attention from bulk mobile-hydrogen abundance towards the local state of defect-trapped hydrogen as a potentially decisive factor in HE. Together with recent advances in quantifying deuterium trapped at dislocations in ferritic steels[12,13], this perspective may enable more quantitative investigations of the relationship between local hydrogen occupancy and fracture susceptibility. It may also help rationalize the narrow temperature window commonly observed for HE[41], given the sensitivity of weakly bound hydrogen at dislocations[42] to its local retention conditions, and is consistent with recent observations by Takai[43,44] of embrittlement associated with weakly bound hydrogen. In contrast to atomistic frameworks that couple HIC nucleation with hydrogen-modified dislocation mobility[29,30,33,45,46], our results support an experimentally testable mechanism in which DTH-dependent core reconstruction facilitates primary atomic decohesion, while recombination-associated energy release may contribute to subsequent crack development. Quantifying the local DTH state may therefore provide a route towards a more predictive, defect-specific understanding of hydrogen-induced fracture.

**Acknowledgments:** Fruitful discussions with Xufei Fang on hydrogen-dislocation interaction, with Daniel Mason on implantation-induced stress evaluation, and with Armin Manhard on the quantitative evaluation of screw dislocation segments are highly appreciated. Thanks are further due to Michael Fußeder and Joachim Dorner for help with the tungsten ion implantations and nuclear reaction analysis, and to Katja Hunger for sample preparation and SEM characterization. During the preparation of this manuscript, the authors used ChatGPT to polish the language and improve readability. All authors reviewed and edited the content as needed and take full responsibility for the final version of the manuscript.

**Funding:** This work has been carried out within the framework of the EUROfusion Consortium, funded by the European Union via the Euratom Research and Training Programme (Grant Agreement No 101052200 - EUROfusion). Views and opinions expressed are however those of the author(s) only and do not necessarily reflect those of the European Union or the European Commission. Neither the European Union nor the European Commission can be held responsible for them.

**Author contributions:**

Conceptualization: LGO

Methodology: LGO, TSS, WJB, MBN, CLI

Investigation: LGO, MBN, CLI

Visualization: LGO, CLI, PMZ, JWG, GHL

Funding acquisition: WJB, RNU, CLR

Project administration: LGO, WJB, RNU, CLR

Supervision: WJB, CLR

Writing – original draft: LGO, PMZ, TSS

Writing – review & editing: all

**Competing interests:** Authors declare that they have no competing interests.

**Data and materials availability:** All data are available in the main text or the supplementary information.

# Supplementary Information

# for

## Defect-trapped hydrogen couples stress-triggered atomic decohesion with explosive crack growth

L. Gao[1,2]*†, T. Schwarz-Selinger[2], M. Balden[2], C. Li[3], P. Manz[4], W. Jacob[2], R. Neu[2], Ch. Linsmeier[1], J. Wang[5]* and G.-H. Lu[6]*

**Affiliations:**

[1]Forschungszentrum Jülich GmbH, Institute of Fusion energy and Nuclear waste management – Plasmaphysik (IFN-1), 52428, Jülich, Germany;
[2]Max-Planck-Institut für Plasmaphysik (IPP), 85748, Garching, Germany;
[3]Shanghai Institute of Applied Physics, Chinese Academy of Science, 201800, Shanghai, China;
[4]University of Greifswald, Institute of Physics, 17489, Greifswald, Germany;
[5]Department of Mechanical and Materials Engineering, University of Nebraska, Lincoln 68588, USA
[6]Beihang University, School of Physics, 100191, Beijing, China.

*Corresponding authors: li.gao@extern.fz-juelich.de (L. Gao), jianwang@unl.edu (J. Wang), lgh@buaa.edu.cn (G.-H. Lu)
†Present address: Gauss Fusion GmbH, 85748, Garching, Germany

**The SI includes:**

## Supplementary Information

## Methods

Polycrystalline, hot-rolled tungsten (W) samples with 99.97 wt.% purity (Plansee SE Austria) and dimensions of $5\times10\times0.8$ mm$^3$ were mechanically polished to a mirror finish[47] and then electro-chemical-polished in a 1.5 *wt.*% NaOH solution. After polishing, all samples were annealed/recrystallized at 2000 K for 30 min in vacuum to reduce the defect density and desorb possible gaseous inclusions. Because recrystallization is performed after polishing, surface distortions introduced by the polishing procedure are also annealed. After recrystallization, the dislocation density turns out to decrease by two orders of magnitude compared with that in the as-delivered sample[22] ($\sim2\times10^{14}$ m/m$^3$). The surviving dislocations in the recrystallized W samples are supposed to be dominated by screw dislocations. The grain size in the recrystallized W sample ranges from 10 to 50 µm as observed by scanning electron microscopy (SEM). The recrystallized W samples were then exposed to deuterium (D) plasma for D decoration of intrinsic defects to a certain occupancy level.

Deuterium plasma exposures were performed in a well-quantified electron-cyclotron-resonance plasma source, housed at Max-Planck-Institute for Plasma Physics (IPP), Garching, Germany. At an operating pressure of 1.0 Pa, an ion flux of primarily $D_3^+$ ions (94%) with minor contributions of $D_2^+$ (3 %) and $D^+$ (3 %) is delivered[48]. The D ion flux at floating potential (~15 eV) was $\sim5.8\times10^{19}$ D m$^{-2}$s$^{-1}$ with an energy of 5 eV per deuteron for the dominant ion species ($D_3^+$), which was selected for the 1$^{st}$ plasma exposure to a fluence of $1.5\times10^{24}$ D m$^{-2}$. The sample temperature was stabilized at 300 K using an ethanol cooling circuit with a thermocouple integrated in the substrate holder to monitor the sample temperature. Half of one sample (W1) was covered by a W foil during the 1$^{st}$ D plasma exposure. This defines two different areas with and without deuterium decoration of the intrinsic defects. The selected plasma exposure parameters were identical to those in Ref.*[23]* (i.e., 300 K, 5 eV/D from $D_3^+$ ions) but with lower D fluence for the present work ($1.5\times10^{24}$ D m$^{-2}$). The lower D fluence ensures that only existing defects were decorated without any blistering/crack formation in the exposed half. Additional samples (Samples W4 & W5, see Table 1 in the main text) subjected to the 1st D plasma exposure were reserved for thermal de-trapping experiments. After the 1st D plasma exposure, these two samples were selectively heated to desorb trapped D at 370 and 450 K for 3 h in vacuum in order to vary the hydrogen occupancy levels at defects. Previous experience with the thermal desorption of D from the same materials has shown that desorption starts at temperatures just above 300 K – the sample temperature during exposures[49]. We also performed the 1st D plasma exposure for two additional W samples at elevated temperatures (i.e., Sample W2@370 K and W3@450 K, see Table 1). This is to produce samples with different D occupancy levels at dislocations for comparison with samples exposed at 300 K with no post-exposure annealing (the exposed half of sample W1), and samples that received post-exposure annealing at 370 K and 450 K (W4 and W5, see Table 1).

Next, both halves of sample W1 were irradiated with 5 MeV $W^{3+}$ ions ($2.2\times10^{16}$ W $m^{-2}s^{-1}$), as performed in the TOF beamline of the 3 MV tandetron accelerator[50] at IPP Garching, Germany. W ions were created with a cesium sputter source from a tungsten carbide target. The samples were clamped with a molybdenum mask on a water-cooled copper substrate holder. The W beam is focused onto the target position with an electrostatic quadrupole triplet lens for a circular final shape. The $W^{3+}$ beam had roughly a 2 mm width at half maximum intensity[50]. To ensure a comprehensive comparison, all W samples (i.e., A3&A4 in sample W1 and W2-W5) were irradiated with the same $W^{3+}$ beam to an ion fluence of $\sim2\times10^{19}$ W $m^{-2}$. The implantation-induced stress during W ion irradiation is taken as external applied load that triggers crack nucleation at dislocations after D decoration with sufficient occupancy level within the irradiated surface areas.

Finally, a 2nd D plasma loading was performed to inflate the nucleated cavities (see Fig. 1) in the exposed half of sample W1 after the 1st D plasma exposure and W ion implantation. The 2nd D plasma loading after the W self-ion implantation was performed at a temperature of 370 K with a bias voltage of -10 V (i.e., leading to an ion energy of 25 eV). Together with the plasma potential this bias voltage yields a mean energy of 8 eV per D for the dominant ion species $D_3^+$. The target D fluence was $3\times10^{25}$ D $m^{-2}$ to ensure sufficient growth/inflation of the nucleated cracks so that they are detectable in the subsequent SEM characterization. The additional W samples (W2-W5, see Table 1) were either exposed to the 1st D plasma at sample temperature of 370 and 450 K (W2, W3) or plasma exposed at 300 K followed by degassing at 370 and 450 K (W4, W5). The W ion irradiation and the 2$^{nd}$ D plasma exposure were performed under the same conditions as applied for sample W1. Note that no W-foil cover was applied during the 2$^{nd}$ D plasma exposure, i.e., the entire surfaces of samples (i.e., W1-W5) were exposed simultaneously to the 2$^{nd}$ D plasma.

The SEM analyses before and after each plasma exposure were performed using a FEI HELIOS NanoLab 600 Dual Beam microscope, which contains a high-resolution scanning electron microscope and focused ion beam (FIB), both with a maximum acceleration voltage of 30 kV. To allow the characterization of the identical locations before and after each treatment, T-shaped markers were engraved in the center of all samples by FIB cutting. By using an in-lens secondary electron (SE) detector as well as a retractable, segmented concentric backscatter (CBS) detector, we were able to observe the details of the intra-granular cracks at the surface. Inflated cavities were only visible by their induced distortion fields in CBS mode (see also Ref.[23]). An electron energy of 5 keV was chosen during the characterization of the regions of interest from all the samples.

## Supplementary Notes

### Note 1: Statistics of HIC detection

To acquire sufficient statistics during the detection of intra-granular HIC formation using SEM, we analyzed squares of 1.5×1.5 $mm^2$ from all four different areas from sample W1 (i.e., A1-A4 in Fig. 1), as well as from samples W2-W5. In the square of W1_A4 we found 130 in-grain blisters (as in Fig. 2) in Fig. S1 (HIC-corresponding blisters are marked by red circles), while, in the other three areas (W1_A1~A3) and four samples with lower DTH occupancy (W2–W5), less than 5 similar objects were detected. Further analysis revealed that the suspected objects in these areas are actually small dust particles on the sample surface (magenta circles in Fig. S1b). We therefore conclude that the intra-granular HIC formation occurs only in W1_A4, which exhibits a sufficient DTH occupancy level at dislocations before stress application via W ion irradiation and the subsequent 2$^{nd}$ D plasma exposure.

### Note 2 Evaluation of diffusive D due to kinetic de-trapping upon W irradiation

Within the implantation zone, D trapped at defects after the 1$^{st}$ D plasma exposure will be kinetically de-trapped during W ion bombardments[51], building-up a distribution of lattice-dissolved D within the W ion implantation zone. To evaluate this D concentration we first attempted to measure the retained D concentration after the 1$^{st}$ D plasma exposure using *D* ($^3$*He*, $p_0$) $\alpha$ nuclear reaction analysis (NRA)[52]. However, due to the extremely low defect density in the recrystallized W materials[22], the concentration of retained D in the samples after the 1$^{st}$ D loading turned out to be below the NRA detection limit (~$10^{-5}$ D/W). This is at least three orders of magnitude lower than that in W samples pre-damaged with 20 MeV W ions and then exposed to D plasma with comparable irradiation conditions[51], where ~2% D/W concentration was detected throughout the whole damaged layer. In Ref.[51], the authors performed an SDTrimSP simulation[53] to evaluate the solute D concentration due to kinetic de-trapping during W ion irradiation. The acquired solute D concentration in equilibrium ranged from $10^{-9}$ to $10^{-8}$ D/W[51]. If we take the present $10^{-5}$ D concentration as the initial value can make a linear extrapolation from Ref. [51] with the ~2 at. % initial D concentration, the solute D concentration in equilibrium for the present experiment during the applied 5 MeV W ion irradiations is $10^{-12}$ to $10^{-11}$ D/W. Such a low solute D concentration is therefore expected to make a negligible contribution to crack nucleation under the present conditions. Furthermore, solute D atoms generated by kinetic de-trapping would be quickly retrapped due to the high defect density created by 5 MeV W ions[51]. The solute D atoms are limited within the implantation zone of W ions. The influence of solute D on crack nucleation from dislocations beyond the implantation zone is negligible and, therefore, neglected here.

**Note 3: Justification of screw dislocation segments as HIC nucleation sites**

For the type of defect being responsible for HIC nucleation, grain boundaries can be excluded, since only intra-granular HICs were observed. We further exclude W interstitials since they hardly survive recrystallization[22] (see ***Materials and Methods***). Among the remaining candidates, vacancies and their clusters exhibit typically rather high hydrogen de-trapping energy[54,55] compared with dislocations[22,56]. The lack of HIC formation by just a heat treatment at 370 K, corresponding to a lower DTH occupancy level at defect centers, shall exclude the possibility that vacancies are responsible for HIC nucleation. Although the de-trapping energy has been demonstrated to be dependent on the local hydrogen occupancy level[57], hydrogen de-trapping from vacancies may still occur upon high occupancy level. Nevertheless, since the fracture mechanics of metals are undoubtedly correlated with dislocations rather than vacancies, we, therefore, propose screw dislocations as the responsible crack nucleation sites, as also supported by our recent experiments about cracking of an electron-transparent W sample by D plasma exposure[36]. Arguments against crack nucleation from vacancy clusters or nano-bubbles can also be found elsewhere[58], where the extremely high stress required for nano-bubble formation is almost impossible to achieve by ion irradiation as performed for this work.

**Note 4: Quantitative evaluation of the applied load via W ion implantation**

We have proposed that the threshold decohesion stress ($\sigma_C$) of hydrogen-decorated dislocations depends on the local DTH occupancy level at the segment cores, as illustrated in Fig. 3. We now estimate the tensile strain just beyond the W ion penetration depth generated due to the constrain from the unirradiated bulk. A very recent study[26] has demonstrated that such strain exists in the sample even 6 months after W ion irradiations. Although *in situ* measuring the implantation-induced stress during W ion irradiation is still challenging, it is reasonable for us to take the measured strain after W ion implantation as the lower limit of that during irradiation, i.e., after some relaxation. Following this idea, we try to estimate the implantation-induced strain for our experimental conditions. According to Fig. 1 in Ref. [26] we take $3\times10^{-4}$ as the value for the here-applied irradiation conditions (5 MeV W ion beam damage to ~5 dpa). The fact that HICs can be nucleated under implantation-induced strain on the order of $10^{-4}$ speaks by itself for the occurrence of hydrogen embrittlement and in turn the indispensable role of trapped hydrogen at dislocations.

Using the estimated strain during W-ion implantation, we estimate the stress from $\sigma_{zz} = (K + 4G/3)\,\varepsilon_{zz}$, taking 310 GPa as the bulk modulus (K) and 161 GPa as the shear modulus (G) for tungsten materials at room temperature. We then reach an estimated compressive stress of ~157 MPa within the ion penetration depth. Importantly, such estimated stress is the maximum value at the end of the W ion implantation zone due to the accumulative effect of the bulk stress on the generated defects density[25]. Accordingly, the tensile stress responsible for the initiation of crack nucleation under the applied ion irradiation is estimated to be approximately 157 MPa beyond the W-ion implantation zone (W-IPZ). This is significantly lower even compared with 0.2% of the yield stress of W materials. We, therefore, conclude that ion implantation-induced stress is insufficient to either mobilize or initiate crack nucleation from H-free dislocations in W materials. However, after hydrogen trapping, the estimated tensile stress from the ion irradiation turned out to be sufficient to reach the critical decohesion threshold for HIC nucleation at some specific screw segment cores. This observation is consistent with a substantial reduction in the local decohesion threshold at sufficiently high DTH occupancy, as schematically illustrated in Fig. 3.

Furthermore, as shown in the precedent *Note 3*, the observed HIC density is far lower than the quantified dislocation density. This discrepancy indicates that most of the H-decorated dislocations are not yet initiated towards crack nucleation under the applied ion implantation-induced stress. It seems that the local stochasticity of screw segments strongly affects the coupling of resolved stress components from the applied load. Thus, HIC nucleation at individual screw segments occurs only when the DTH-enabled normal stress coupling is sufficient to overcome the locally reduced decohesion threshold, as illustrated in Fig. 3.

**Note 5: Estimating the recombination-associated energy-density scale in an incipient crack**

The proposed two-stage chemomechanical model reconciles the apparent paradox between the small mechanical perturbation via W ion irradiation and the extreme theoretical cleavage strength of the tungsten lattice. As proposed in the manuscript, once the DTH occupancy reaches unity ($\theta = 1$), the core undergoes structural instability towards atomic decohesion under the applied stress $\sigma_a$ (i.e., the estimated implantation-induced stress during W ion irradiation (~157 MPa)). We consider the chemical energy released by $H_2/D_2$ formation as a possible contributor to the rapid growth following atomic decohesion. Following this proposal, we estimate the nominal energy-density scale associated with $H_2/D_2$ formation under the assumed geometric confinement conditions.

Because deuterium plasma was used in the present experiments, the proposed molecular recombination process is written as ($2D^0 \rightarrow D_2$). The ground-state dissociation energy of $D_2$ is approximately 4.556 eV per molecule[59], corresponding to an energy scale of ($7.30\times10^{-19}$) J per molecule:

$\boldsymbol{\Delta E_{chem} = \Delta H_{rec}} \approx$ 4.556 eV/molecule (439.5 kJ/mol) as the recombination enthalpy.

At the instant of decohesion, if we assume such bond rupture creating an initial atomic gap of width $\boldsymbol{d}$. For a screw dislocation segment of length $\boldsymbol{L}$ with Burgers vector $\boldsymbol{b}$, the infinitesimal volume $\boldsymbol{V_{gap}}$ that confines the $D_2$ molecule can be estimated as $\boldsymbol{V_{gap} \approx b \cdot d \cdot L}$.

Due to the extreme density, the ideal gas law is inapplicable. Rather than solving an equation of state to determine the transient pressure, we estimate the nominal energy density associated with deuterium recombination by dividing the assumed chemical energy released by the effective geometrical volume available to the molecules, i.e., $\boldsymbol{\Delta E_{chem}} / \boldsymbol{V_{eff}}$. To account for the finite excluded volume of the $D_2$ molecule, the effective volume is taken as $\boldsymbol{V_{eff}} = \boldsymbol{V_{gap}} - \mathbf{n}\boldsymbol{\beta}$, where $\boldsymbol{V_{gap}}$ is the assumed geometrical confinement volume, $\mathbf{n}$ is the molar amount of $D_2$, and $\boldsymbol{\beta} \approx 1.55\times10^{-5}$ $m^3$/mol is the molar excluded-volume parameter for deuterium. The nominal energy density is then estimated as $\boldsymbol{u_{nominal}} = \boldsymbol{\Delta E_{chem}} / (\boldsymbol{V_{gap}} - \mathbf{n}\boldsymbol{\beta})$, where $\boldsymbol{\Delta E_{chem}}$ is the assumed chemical energy released by $D_2$ formation.

We further adopt the following experimental parameters for the W system:

$\boldsymbol{b}$ = 0.274 nm as the Burgers vector of ½<111> screw dislocation

$\boldsymbol{d} \approx$ 0.2 nm (approximate atomic decohesion gap)

$\boldsymbol{\theta}$ = 1 (proposed dynamic saturation regime, see Fig. 3).

For one $D_2$ molecule ($\boldsymbol{n} = 1/\boldsymbol{N_A} = 1.7 \times 10^{-24}$ mol , 2 D atoms out of the trapped D cluster) formation within a unit length of a screw segment ($\boldsymbol{L}$= 1 nm), the chemical energy $\boldsymbol{\Delta E_{chem}} \approx \boldsymbol{\Delta H_{rec}} \cdot \boldsymbol{n} = 7.3 \times 10^{-19}$ J, and the excluded volume $\mathbf{n}\boldsymbol{\beta} = 2.6 \times 10^{-29}$ m$^3$, and the available volume $\boldsymbol{V_{gap}} \approx 5. \times 10^{-29}$ m$^3$. The corresponding energy-density scale is therefore $\boldsymbol{\Delta E_{chem}} / (\boldsymbol{V_{gap}} - \mathbf{n}\boldsymbol{\beta}) \approx 2.51 \times 10^{10}$ Pa, approximately 25 GPa.

The estimated value of approximately 25 GPa represents an order-of-magnitude nominal energy-density scale based on the assumed molecular formation energy and geometric confinement volume; it is not a prediction of the transient pressure hence it cannot be directly compared with the ideal cleavage strength of bcc W (~29 GPa[37]) or a significantly lower value at dislocation core compared with perfect lattice[31] to establish a fracture criterion, because the transient pressure and the conversion of chemical energy into mechanical work cannot be quantified without a validated description of the confined molecular state and the associated energy partitioning. In parallel, the formation of internal mesoscale cavities with diameters of several hundred nanometers demonstrates that substantial crack growth occurs under the present experimental conditions without an additional externally applied mechanical load. We therefore treat molecular recombination as a plausible proposal that contributes to rapid crack growth, rather than as a directly established source of a specified pressure. We further refer to the proposed transition from atomic decohesion to rapid crack growth as an “atomic explosion”. This term denotes the hypothesized contribution of recombination-associated energy release to subsequent crack growth; it does not imply that the transient pressure or the fraction of chemical energy converted into mechanical work has been established quantitatively.

**Note 6: HIC nucleation occurs at depth beyond the W ion penetration depth**

During W-ion implantation, the incident ions and the resulting interstitial W atoms produce volumetric swelling and additional strain within the implantation zone[26]. We propose that HIC nucleation is favoured beyond this zone, where tensile irradiation-induced stress is expected and where pre-existing dislocation structures may retain D more stably. Within the implantation zone, irradiation damage and the associated compressive stress are expected to make the proposed nucleation process much less favourable. The available cross-sectional observations are consistent with this interpretation but do not directly establish the original nucleation point.

To demonstrate the postulation that no crack nucleation occurs within the ion implantation zone, we performed a SRIM simulation [60,61] to calculate first the implantation depth of 5 MeV W ions into W matrix. As shown in Fig. S2, the implantation zone extends to a depth of ~0.75 μm. Next, we evaluate the diffusion length of D during the 1st D implantation based on the diffusion coefficient determined by Fraunfelder [62]. This gives rise to a diffusion length of 64.8 μm during ~7 hours plasma exposure at 300 K, which is far larger than the implantation depth of W ions. This is sufficient to ensure potential HIC nucleation at D-decorated dislocations located from 0.75 to ~65 μm. Within the implantation zone, the damage level of tungsten matrix is shown to be more than 5 dpa (i.e., >> 0.1 dpa, the damage level corresponding to the transition of the strain from compressive to tensile[26]), which ensures a compressive strain within- and tensile strain beyond the implantation zone in the order of $10^{-4}$ (see Fig. 1 in Ref. [26]). Finally, using the focused ion beam integrated in our SEM we conducted cross-sectioning of several intra-granular cracks after detecting them as in-grain blisters. One such example is shown in Fig. 2 in the main text: the surface morphology before cutting (Fig. 2B&C1) and the intra-granular crack beneath the surface (Fig. 2-C2). The cavity midpoint at approximately 850 nm is consistent with nucleation beyond the estimated W-ion penetration depth, although the original nucleation point cannot be determined directly from the final cavity geometry. Cross-sections of other in-grain blisters also support this conclusion. This observation that the cavity midpoint lies beyond the estimated maximum W-ion implantation depth is consistent with the proposed nucleation of HICs beyond the implantation zone, where tensile irradiation-induced stress is expected. However, the cross-sectional geometry does not directly reveal the original nucleation point, and the inferred location should therefore be regarded as consistent with, rather than definitive proof of, the proposed mechanism. Interestingly, we also found a few nanopores in isolated W grains during cross-sectioning, but such nanopores are believed to be irrelevant to the nucleation of intra-granular cracks, since the required gas pressure inside the pore to initiate crack nucleation (>1 GPa [58]) is far beyond that expected under the applied irradiation conditions.

In this context, it is worth noting that the strategy applying external load via self-ion implantation is not applicable to be combined with *in situ* TEM techniques for direct crack nucleation observation, because TEM techniques are limited to electron-transparent samples with thickness of roughly 100 nm for W. This is only 1/7$^{th}$ of the ion penetration depth of 5 MeV W ions. As a result, the here-applied W ion implantation generating appropriate stress for decohesion cannot be implemented for initiating HIC nucleation in electron-transparent W samples.

**Note 7: Quantitative evaluation of screw dislocation segments**

Detailed characterization of dislocations surviving after W sample recrystallization at 2000 K can be found in Ref.[22,49]. Representative screw dislocation segments in the TEM-examined area of the recrystallized sample have an average length of ~100 nm. The density of such screw segments[22] was evaluated $(2\pm1.4) \times 10^{12}$ m/m$^3$, being two orders of magnitude with respect to the initial dislocation density in the as-delivered W samples. Assuming that these 100 nm long screw segments are the responsible HIC nucleation sites for the observed blisters (Fig. 2), we have estimated the number density of in-grain blisters in W1_A4 with the complete sequential irradiations.

Note that only dislocations located beyond the penetration depth of 5 MeV W ions (i.e., 750 nm, see Fig. S2) can hold hydrogen stably and in turn respond to the ion-implantation-induced stress generated beyond the W ion penetration depth. We, therefore, take the depth range from 750 to 1600 nm as the potential crack-relevant depth for the examined area shown in Fig. S1 (1.5×1.5 mm$^2$). Assuming all dislocations in this volume (1.5 mm×1.5 mm× 0.85 μm) could act as crack-nucleation sites after reaching local DTH saturation to a certain occupancy level, however, gives rise to more than $10^7$ blisters in the examined area. The fact that only 130 blisters were finally observed strongly points to the invalidity of the assumption taking all ~100 nm dislocations segments as HIC nucleation sites. Obviously, the HIC nucleation process is strongly depending on the local configuration of the dislocation segments and supposed to be a rare event under the applied conditions. Specific qualifications for dislocations to be triggered as HIC nucleation sites include the localized DTH occupancy level and also the geometry of the dislocation towards effective coupling with the resolved normal stress (see Fig. 3) under the applied conditions.

**Note 8: Estimating the averaged DTH line density at screw segments**

Since we observed HIC nucleation in W samples with the 1st D plasma exposure at 300 K but not at 370 K, it is reasonable to take for the corresponding DTH occupancy level ranging between these two conditions based on the D concentration profiles in W after D plasma exposure at 300 K ($2.5\times10^{-3}$ at.%[23]) and 370 K ($1.8\times10^{-3}$ at.%, see D depth profile determined by nuclear reaction analysis (NRA) in Fig. S3). We assume negligible contribution to D retention from vacancies or impurities in the high-purity recrystallized W sample. Also, for the depth relevant to in-grain HICs (750-1600 nm, see **Fig. S3**) we neglect the retained D at grain boundaries in the recrystallized W sample and attribute all of the retained D to trapping at dislocations with a density[22,49] of $(2\pm1.4) \times 10^{12}$ m/m$^3$. This yields an estimate of the averaged DTH line density at dislocations in the recrystallized W sample: $3.5\text{-}26.3\times10^5$ D atoms per meter dislocation line. Notably, as contributions from vacancies and impurities are neglected, the estimate of averaged DTH line density is supposed to represent an upper limit in terms of the effective DTH occupancy level ensuring the initiation of crack nucleation by the applied W ion irradiation-induced stress. Compared with the required local DTH saturation, it seems that the average DTH line density is far too small. However, it is known that DTH atoms will not be uniformly distributed along dislocation lines. At some screw segments, the local DTH occupancy level that ultimately enables the final HIC nucleation is supposed to be significantly higher than the averaged DTH line density estimated above. In other words, the estimated low DTH line density does not deviate from the proposed 2-stage mechanochemical model on HIC nucleation (see Fig. 3&4).

***Supplementary Figure S1***

***Statistical SEM survey of HIC-associated features.*** *Backscattered-electron images of 1.5 × 1.5 mm² areas: a) W1_A4, showing HIC-associated blisters (red circles); b) W1_A3, where no HICs were detected. The same absence was observed in W1_A1 and W1_A2. The dashed box in a) marks the area shown in Fig. 2A. Magenta circles indicate surface dust particles.*

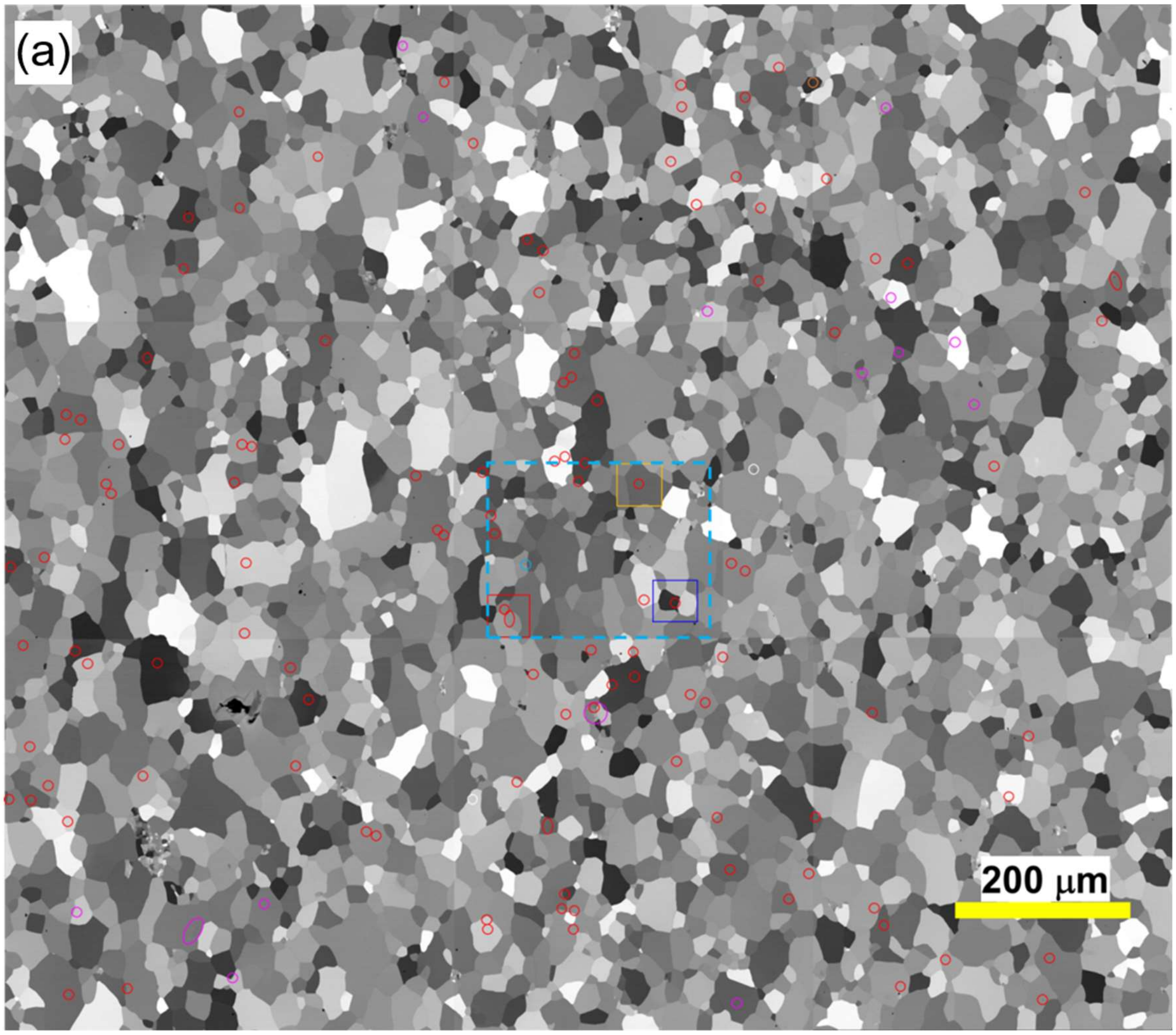

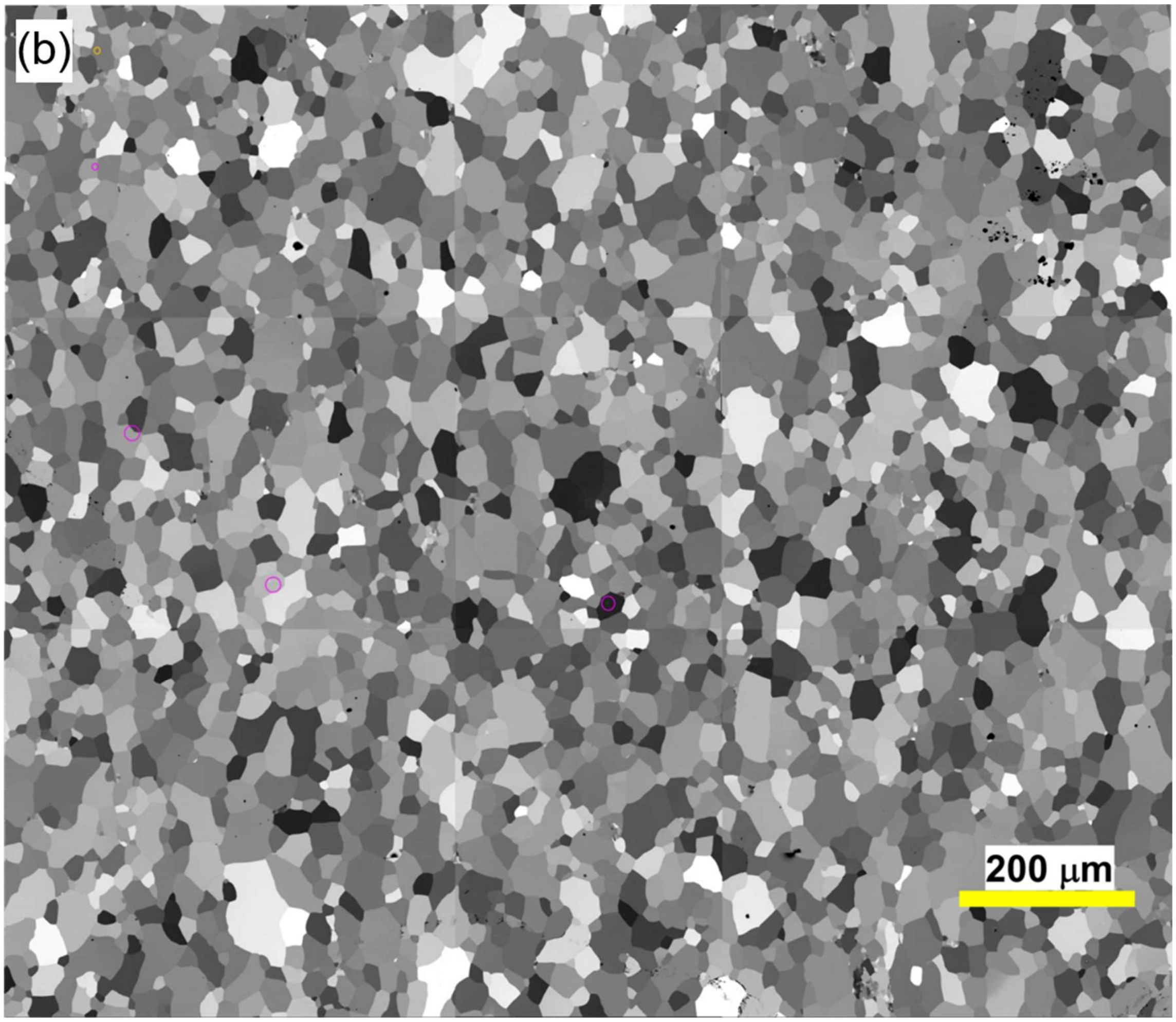
(b)
200 μm

***Supplementary Figure S2***

***SRIM simulation of 5 MeV W ion implantation into W sample.*** *The implanted-ion distribution (blue) and damage profile (red) are shown for a displacement threshold energy of 90 eV, as recommended by the American Society for Testing and Materials[56]. Shown are the distribution of implanted W ions (in blue) and the damage profile (in red). In this case, the largest penetration depth of implanted W ions is about 0.75 μm. The W ion fluence was $2\times10^{19}$ W/m$^2$ in the center of the target position for the experiments performed in this work.*

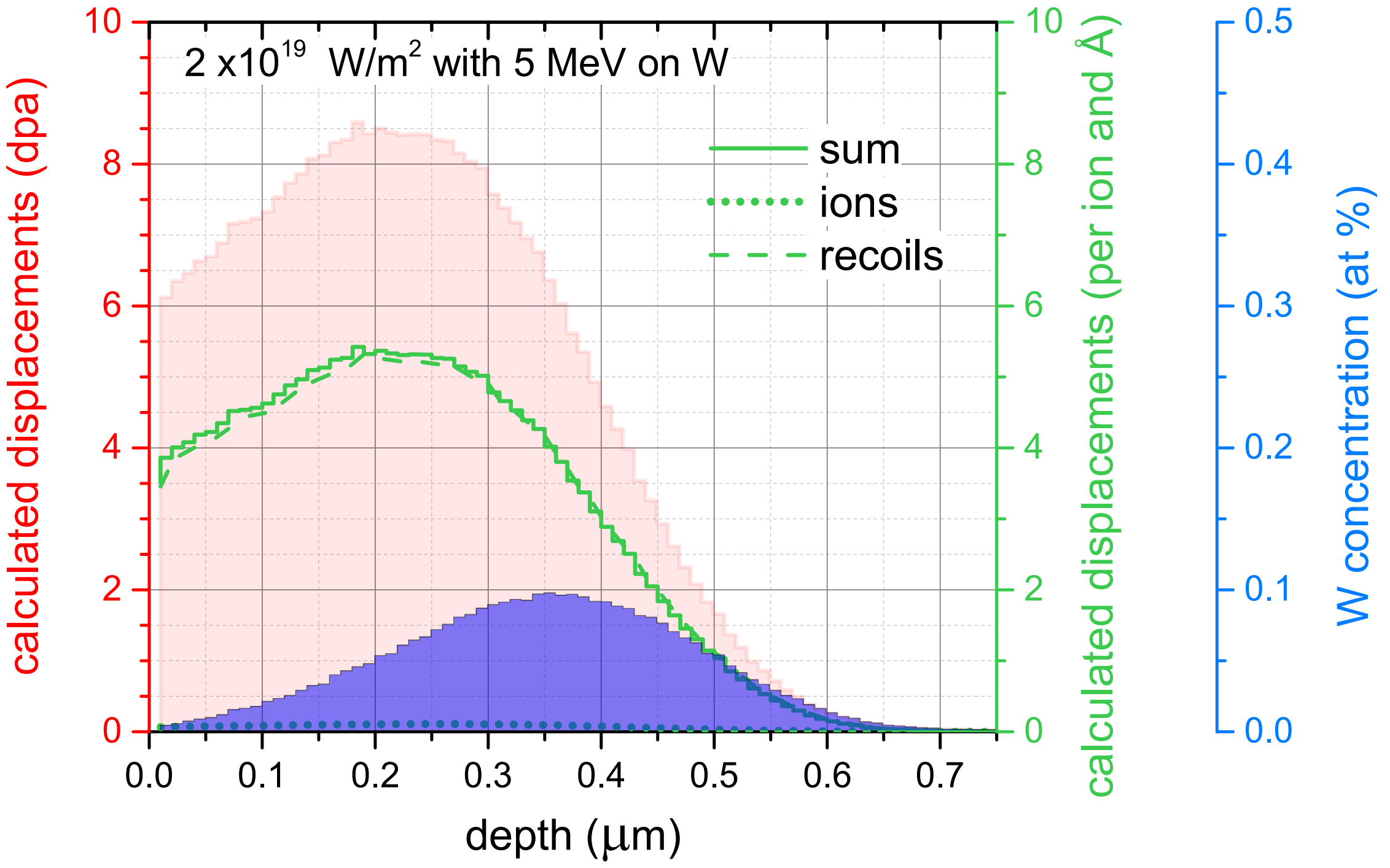

***Supplementary Figure S3***

***Deuterium depth profile in recrystallized tungsten.*** *The sample was exposed to D plasma at 370 K with floating potential to a fluence of* $1\times 10^{25}$ *D/m*$^2$*. The retained D concentration, measured by* $^3$*He nuclear reaction analysis, was* $1.8\times10^{-3}$ *at. %, close to the detection limit under these conditions.*

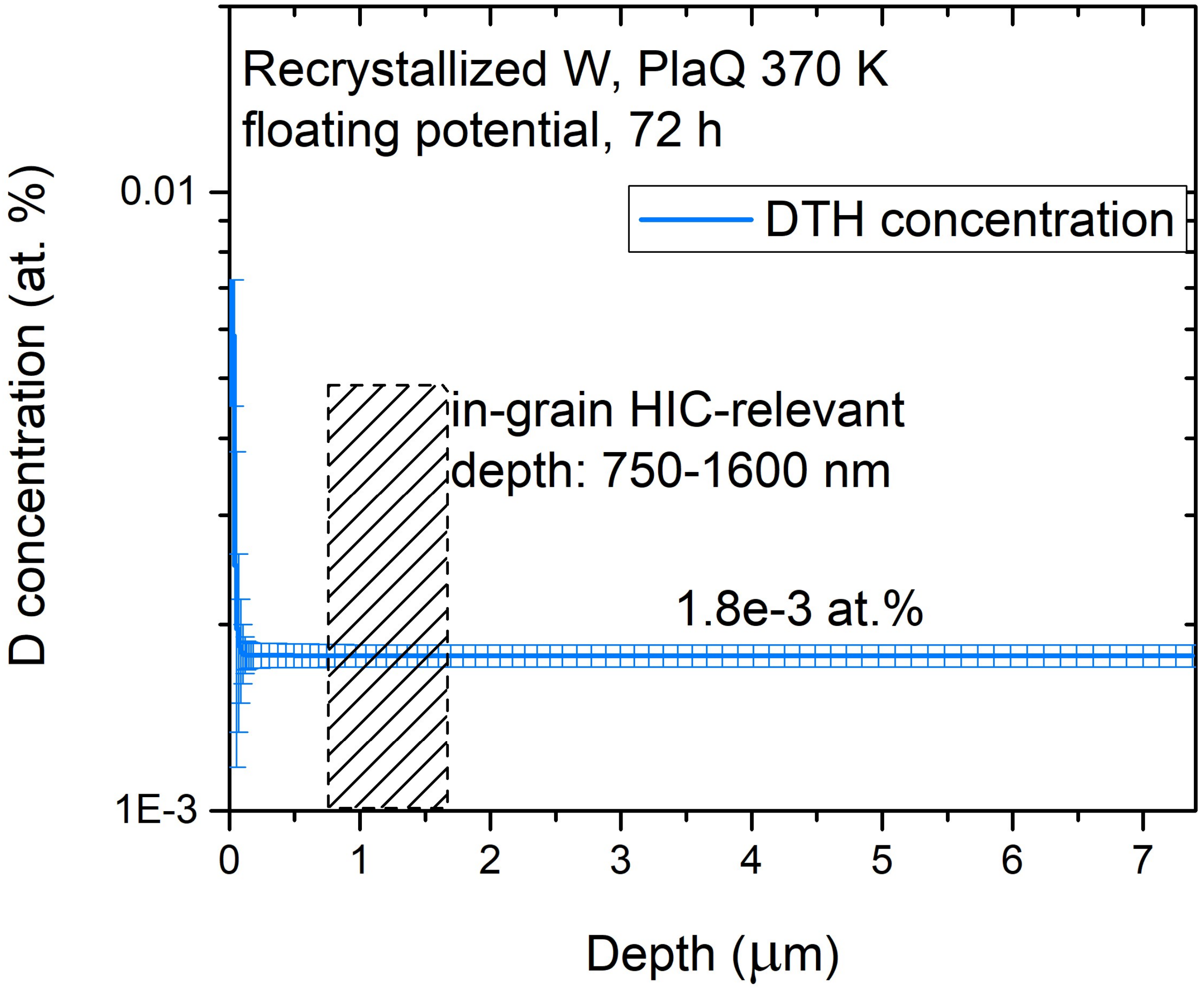